\documentclass[12pt]{article}
\usepackage[utf8]{inputenc}
\usepackage{amssymb,amsmath,amsthm}
\usepackage{mathtools}
\usepackage{bbm}
\usepackage{empheq}
\usepackage{graphicx}
\usepackage[normalem]{ulem}
\usepackage{enumitem} 

\DeclareMathOperator{\diag}{diag}
\usepackage{booktabs} 

\usepackage[english]{babel}
\usepackage[T1]{fontenc}
\usepackage[dvipsnames]{xcolor}
\usepackage{appendix} 
\usepackage{bm}
\usepackage{lipsum}
\usepackage{hyperref}
\usepackage{graphicx}
\usepackage{siunitx}
\hypersetup{colorlinks,linkcolor={blue},citecolor={blue},urlcolor={blue}} 
\usepackage[authoryear]{natbib}
\usepackage{gensymb}
\usepackage{titlepic}
\usepackage{array,ragged2e}
\usepackage{rotating}
\usepackage{tikz}

\numberwithin{equation}{section}
\theoremstyle{plain}

\theoremstyle{remark}

\theoremstyle{remark}

\usepackage{colortbl}
\usepackage[dvipsnames]{xcolor}
\usepackage[total={170mm,230mm},
 left=25mm,
 top=20mm]{geometry}
\definecolor{rev1}{RGB}{216,27,96}
\definecolor{rev2}{RGB}{30,136,229}
\definecolor{rev3}{RGB}{181,103,33}

\newgeometry{top=5mm,left=25mm,right=25mm,bottom=25mm} 
\title{Adaptive Bayesian Very Short-Term Wind Power Forecasting Based on the Generalised Logit Transformation}

\author{Tao Shen$^1$*, Jethro Browell$^1$, Daniela Castro-Camilo$^1$}

\footnotetext[1]{
\baselineskip=10pt School of Mathematics and Statistics, University of Glasgow, UK.}

\date{2025, March, 20}

\begin{document}
\maketitle

\baselineskip=16pt
\begin{center}

{\large{\bf Abstract}}
\end{center}
Wind power plays an increasingly significant role in achieving the 2050 Net Zero Strategy.
Despite its rapid growth, its inherent variability presents challenges in forecasting.
Accurately forecasting wind power generation is one key demand for the stable and controllable integration of renewable energy into existing grid operations.
This paper proposes an adaptive method for very short-term forecasting that combines the generalised logit transformation with a Bayesian approach.
The generalised logit transformation processes double-bounded wind power data to an unbounded domain, facilitating the application of Bayesian methods.
A novel adaptive mechanism for updating the transformation shape parameter is introduced to leverage Bayesian updates by recovering a small sample of representative data.
Four adaptive forecasting methods are investigated, evaluating their advantages and limitations through an extensive case study of over 100 wind farms ranging four years in the UK.
The methods are evaluated using the Continuous Ranked Probability Score and we propose the use of functional reliability diagrams to assess calibration.
Results indicate that the proposed Bayesian method with adaptive shape parameter updating outperforms benchmarks, yielding consistent improvements in CRPS and forecast reliability.
The method effectively addresses uncertainty, ensuring robust and accurate probabilistic forecasting which is essential for grid integration and decision-making.

\par\vfill\noindent

\restoregeometry 
\newgeometry{top=20mm,left=25mm,right=20mm,bottom=20mm}

\baselineskip=20pt

\newpage


\section{Introduction}\label{sec:intro}

Wind energy is playing a key role in the energy transition.
In 2023, it contributed $7.8\%$ to power generation globally \citep{ember_global_electricity_review_2024} and installed capacity has increased from $416$GW in 2015 to $1017$GW in 2023 \citep{ourworldindata_wind_capacity}.
Furthermore, at COP28, 130 governments committed to triple renewable energy capacity to 11,000 GW by 2040 \citep{unfccc_gca_summary_cop28}.
These figures highlight the importance and potential of wind power.
However, like other natural resources such as hydropower or solar power, wind power is inherently variable due to its weather-dependent nature.
Factors such as regional climate and local orography contribute to this uncertainty \citep{hanifiCriticalReviewWind2020}.
To ensure a smooth integration of wind energy, effective forecasting methods and regulations are essential for managing electricity supply, demand, and distribution.

A key challenge in wind energy forecasting is Very-Short-Term Forecasting (VSTF), which focuses on lead-times typically less than six hours \citep{Nowcasting}.
These forecasts are updated frequently as new observational data become available.
While the six-hour threshold is not rigid, it marks the point where observations provide the main source of predictability, as opposed to Numerical Weather Prediction (NWP; \citealp{tawnReviewVeryshortterm2022,zhangReviewProbabilisticforecasting2014}), which is constrained by latency associated with data assimilation and computation \citep{BLAGA2019119,tawnReviewVeryshortterm2022}. 
In contrast, statistical methods, which leverage near real-time observations, typically yield more accurate forecasts. 
Even with the advances in AI-driven weather models that significantly reduced computational time, the latency of data retrieval from global observation networks means statistical methods are likely to remain superior for forecast horizons of minutes to a few hours \citep{TheImpactofDataLatencyonOperationalGlobalWeatherForecasting}.

Classical time series methods have been widely applied to forecast wind power generation directly from past data.
Models such as Auto-Regressive (AR), Moving Average (MA), and Autoregressive Integrated Moving Average (ARIMA) \citep{TORRES200565,pmdarima} have been widely discussed in the literature \citep{PETROPOULOS2022705,hanifiCriticalReviewWind2020}.
Vector Auto-Regressive models have been used to exploit spatio-temporal dependency to improve forecast performance, though the high-dimension of the problem represents a challenge, which has been addressed using LASSO, amongst other methods \citep{dowellVeryShortTermProbabilisticWind2015,cavalcanteLASSOVectorautoregression2017,messnerOnlineAdaptivelasso2019}.
Furthermore, recursive updates---also called adaptive or online updates---can be performed to track non-stationarities and improve performance empirically \citep{aitmaatallahRecursiveWindspeed2015,messnerOnlineAdaptivelasso2019}.
For instance, the Recursive Least Squares algorithm provides an online update based on Ordinary Least Squares estimation \citep{pinsonVeryShortTermProbabilisticForecasting2012}, and online estimation of forecast error variance using exponential smoothing was proposed in \cite{dowellVeryShortTermProbabilisticWind2015}.

Beyond statistical models, effective data preprocessing can further enhance forecasting performance. 
This is particularly important for wind power data, which is double-bounded—--restricted between zero and its rated power—--and subject to operational interventions that modify the time series characteristics.
A common approach for handling one-sided bounded data is the Box-Cox transformation \citep{boxcox1964}.
Additionally, the Logit-Normal distribution, introduced by \cite{aitchisonLogisticNormalDistributionsProperties1980}, provides a useful framework for modelling such data.
Furthermore, \cite{pinsonVeryShortTermProbabilisticForecasting2012} and \cite{Amandine2021} demonstrate that forecasting performance can be improved using a generalised logit transformation.

Although deterministic forecasting was considered sufficient practice in the 1990s, a shift towards probabilistic forecasting gradually emerged with the work of, e.g., \cite{Gneiting2008}.
Nowadays, decision-makers can benefit from more information to quantify the risk and impact of uncertainty \citep{dowellVeryShortTermProbabilisticWind2015}.
One approach to generating probabilistic forecasts is through ensemble methods, which involve running multiple simulations with varying initial conditions \citep{Buizza2008}
To assess the quality of probabilistic forecasting, proper scoring rules such as the Brier score \citep{brierVERIFICATIONFORECASTSEXPRESSED1950} and the continuous ranking probability score (CRPS; \citealt{Gneiting01032007,gneitingProbabilisticForecasting2014}) have been proposed.

Bayesian methods are particularly well-suited for probabilistic forecasting as they are inherently probabilistic, and various studies have demonstrated their effectiveness across different applications.
For instance, \cite{garbatov2002bayesian} introduced a Bayesian framework to iteratively update reliability estimates for floating marine structures, integrating inspection and maintenance data to refine probabilistic models.
In wind speed forecasting, \cite{bracaleAdvancedBayesianMethod2015} used a mixture of two Weibull distributions to quantify uncertainty, while \cite{dccsplicedGammaGP2019} proposed a spliced model based on extreme value theory to capture both extreme and non-extreme wind speeds.
Bayesian methods have also been successfully integrated with machine learning techniques.
\cite{blonbouVeryShorttermwind2011} and \cite{GARCIA202055} demonstrated their flexibility by combining them with neural networks for wind speed and wind power forecasting. 
Further advancing this approach,
\cite{TANG2022e11599} proposes an innovative Bayesian ensembling model that enhances wind power prediction accuracy by combining multiple forecasting methods to better account for inherent uncertainties. 
However, with the exception of \cite{dccsplicedGammaGP2019}, most of these studies focus primarily on deterministic forecasting and associated evaluation metrics, such as mean squared error or mean absolute error, rather than evaluating the probabilistic forecasts that their methods produce implicitly.

Here, we propose a method for probabilistic forecasting of wind power production, integrating the inflated Generalised Logit distribution with a novel adaptive Bayesian inference scheme.
While the logit transformation is widely used, its generalised form has received little attention, to the best of our knowledge.
A key feature of our proposed online Bayesian estimation is a recursive update scheme for the shape parameter of the Generalised Logit distribution, allowing for adaptive learning over time.
to rigorously evaluate our method, we conducted an extensive case study of more than 100 wind farms in Great Britain, comparing our approach to benchmark models using tools from functional data analysis to effectively visualise and interpret the results across all wind farms.
We show for the first time the extent to which the relative performance of different online learning algorithms varies between wind farms, providing new insights into the generalisability (or lack thereof) of such methods in wind power forecasting.
Through comprehensive evaluation and sensitivity analysis, we demonstrate that our adaptive Bayesian estimation not only achieves the highest forecast accuracy but also ensures greater robustness compared to state-of-the-art benchmarks.

The rest of this paper is organised as follows.
Section \ref{sec:modelling} introduces the Generalised Logit distribution, which we employ for data processing, and presents the auto-regressive model used in our wind power generation forecasting method.
Section~\ref{sec:background} reviews classical estimation methods for comparison with our approach.
Section~\ref{sec:ourmodel} outlines our proposed modelling framework.
Section \ref{sec:exp} presents experiments and evaluation results using the collected wind farm power generation data, followed by the conclusion in Section \ref{sec:cls}.

\section{Modelling Approach}\label{sec:modelling}

This section lays the groundwork for adaptive probabilistic forecasting by transforming wind power data into a statistically tractable form before applying time series forecasting methods. 
Wind power data are inherently bounded and often skewed, particularly near zero and rated power, posing challenges for direct statistical modelling. 
To overcome this, we employ the generalised logit transformation, which rescales the data and introduces a shape parameter to account for skewness.
This enhances flexibility and accuracy while only slightly increasing model complexity.
Additionally, we incorporate censoring and thresholding techniques in auto-regressive modelling to effectively handle boundary values, ensuring a more robust and reliable representation for forecasting.

\subsection{Generalised Logit Transformation}\label{sec:def}

Let $P_\text{rated}$ be the rated power (in mega-watts, MW) of a given wind farm, and $\{X_t\}, t = 1,2,3 ...$ the time series of random variables with support $[0, P_\text{rated}]$.
Then $\{X_t / P_\text{rated}\}, t = 1,2,3 ...$ is a re-scaled time series with support $[0, 1]$. 
For a random variable $X$ with support $(0,1)$, the generalised logit transformation is defined as
\begin{equation}\label{eq:Logit} 
    Y := {L_\nu}(X) = \ln \frac{X^\nu}{1-X^\nu} \qquad X \in (0,1),
    \end{equation}
where $\nu \in (0, +\infty)$ is a hyper-parameter called the shape parameter.
The inverse transformation is given by
\begin{equation}\label{eq:inv-logit} 
    X := {L_\nu}^{-1}(Y) = \left(\frac{e^{Y}}{1 + e^{Y}} \right)^{\frac{1}{\nu}} \qquad Y \in (-\infty,+\infty)
    .\end{equation}
The shape parameter $\nu$ plays an important role since it accounts for potential skewness in the data.
As noted by~\cite{dowellVeryShortTermProbabilisticWind2015} and~\cite{pinsonVeryShortTermProbabilisticForecasting2012}, when a wind farm operates at near its rated power or zero power, the conditional distribution of wind power generation is skewed.
Notably, when $\nu=1$, the transformation in \eqref{eq:Logit} simplifies to the standard logit transformation. 
If $Y$ is assumed to follow a normal distribution with mean $\mu$ and variance $\sigma^2$, denoted $Y \sim \mathcal{N}(\mu, \sigma^2)$, then $X$ follows the logit normal distribution \citep{aitchisonLogisticNormalDistributionsProperties1980}, denoted as $\mathcal{L_\nu}(\mu,\sigma^2)$, with probability density function
\begin{equation} \label{eq:Lv Dits} 
    p({x}|\mu,\sigma,\nu) =  
    \frac{1}{\sqrt{2\pi \sigma^2}} 
    \frac{\nu}{x(1-x^\nu)}
    \exp \Bigl\{ -\frac{( {L}_\nu(x) - \mu) ^2}{2\sigma^2} \Bigl\} .\end{equation}

The generalized logit transformation in \eqref{eq:Logit} naturally excludes 0 and 1 from its support.
However, wind farms operate at either zero power---due to low wind conditions or high-wind shutdown---or at their  rated power.
This introduces a boundary issue that must be addressed for effective statistical modelling.
One approach to handling this is censoring \citep{lesaffreLogisticTransformbounded2007}, 
which extends the support of $X$ to $[0, 1]$ by assigning probability mass at the boundaries, leading to the censored probability distribution
\begin{equation}
    X \sim \omega_{0}\delta_{0} + (1-\omega_{0}-\omega_{1})\mathcal{L}_\nu(\mu,\sigma^2) + \omega_{1}\delta_{1} \qquad \omega_{0},\omega_{1} \in [0,1],\qquad 0\leq \omega_{0}+\omega_{1} \leq 1
    ,\end{equation} 
where $\delta_{0}$ and $\delta_{1}$ denote the Dirac delta functions at $0$ and $1$, respectively, and $\omega_{0}$ and $\omega_{1}$ are probability masses allocated at $-\infty$ and $+\infty$, respectively.
Since $Y$ is the generalised logit transformation of $X$, its support extends to $[-\infty, +\infty]$, with probability distribution
\begin{equation}
    Y \sim \omega_{0}\delta_{0}^\prime + (1-\omega_{0}-\omega_{1})\mathcal{N}(\mu,\sigma^2) + \omega_{1}\delta_{1}^\prime \qquad \omega_{0},\omega_{1} \in [0,1],\qquad 0\leq\omega_{0}+\omega_{1}\leq 1
    \end{equation} 
where $\delta_{0}^\prime$ and $\delta_{1}^\prime$ are the Dirac Delta distributions at $-\infty$ and $+\infty$, respectively, and $\omega_{0}=$ and $\omega_{1}$ are probability masses allocated at $-\infty$ and $+\infty$, respectively.

\cite{pinsonVeryShortTermProbabilisticForecasting2012} proposed an alternative solution to the boundary issue based on thresholding the original observed data.
Under this approach, a threshold $\epsilon \in (0,1)$ is chosen based on the measurement precision of the data $X$---typically, a small value such as $\epsilon = 0.005$. 
This results in a transformed support, where $X$ is constrained to $[\epsilon, 1-\epsilon]$ and $Y$ is constrained to $[{L_\nu}(\epsilon), {L_\nu}(1-\epsilon)]$. 
Conceptually, thresholding treats all values in $[0, \epsilon)$ as zero and those in $(1-\epsilon, 1]$ as 1. 
Another interpretation is that values below $\epsilon$ are \emph{squeezed} into the lower bound, while those above $1-\epsilon$ are \emph{squeezed} into the upper bound.
Since each interval is represented by a single value, probability mass is accumulated at the boundaries.
Consequently, $Y$ now follows an inflated normal distribution $\mathcal{N}^\star (\mu, \sigma^2, {L_\nu}(\epsilon), {L_\nu}(1-\epsilon))$ with cumulative density function
\begin{align}
    F(Y) =
    \begin{cases} 
        0 & Y < L_\nu(\epsilon), \\
        \Phi(Y) & L_\nu(\epsilon) \leq Y < L_\nu(1-\epsilon), \\
        1 &  L_\nu(1-\epsilon) \leq Y,
    \end{cases}
\end{align}
where $\Phi(\cdot)$ is the cumulative density function of normal distribution $\mathcal{N}(\mu,\sigma^2)$. 
It is crucial to distinguish an inflated normal distribution from a truncated normal distribution.
Given a normally distributed random variable with probability density function $\varphi(\cdot)$ and cumulative density function $\Phi(\cdot)$,
the corresponding truncated normal distribution with support restricted to $[a,b]$ has probability density function
$p(x) = \varphi(x)/({\Phi(b) - \Phi(a)})$.
Both the truncated and inflated normal distributions have bounded support, but they redistribute probability mass differently. 
The truncated normal distribution redistributes probability mass from $(-\infty,a]$ and $[b,+\infty)$ evenly across the interval $[a,b]$ via the normalisation term $\Phi(b) - \Phi(a)$, while the inflated normal distribution concentrates probability mass at the boundaries $a$ and $b$, without altering the density within $(a,b)$.

\subsection{Auto-Regressive Model for Wind Energy Time Series} \label{sec:model}

From this section onwards, $\{X_t\}$ will denote the stochastic process of wind power generation, re-scaled and bounded to $[\epsilon,1-\epsilon]$ with regard to the rated power and the threshold $\epsilon$. 
and $\{x_t\}$ the corresponding observed time series data.  
As before, let $\{Y_t\}$ be the Logit-Normal transformed process of $\{X_t\}$, with the choice of $\nu$ to be determined.
We use $\{x_t\}$ and $\{y_t\}$ to denote the corresponding transformed observations.
To learn from historical measurement records, we assume that $\{y_t\}$ follows an auto-regressive model of order $p$, AR($p$), i.e.,
\begin{equation}\label{eq:model yt}
    y_t = \theta_0 + \theta_1 y_{t-1} + \theta_2 y_{t-2} + ... + \theta_p y_{t-p} + z_t = \mathbf{y}_{t,B_p}^\top\boldsymbol{\theta} + z_t,
    \end{equation}
where $\boldsymbol{\theta}=[\theta_0, ..., \theta_p]^\top$ is the vector of model parameters, $z_t$ is an error term, 
$B_p$ is the backshift operator \citep{Box2016} and $\mathbf{y}_{t,B_p}$ the vector for $p$ past observations at time $t$, i.e., $\mathbf{y}_{t,B_p} = [1,y_{t-1},\ldots, y_{t-p}]^\top$.
For readability purposes, the rest of the paper will abbreviate the subscript $\cdot_{t,B_p}$ with $\cdot_B$ when there is no confusion to avoid notational clutter.

The parameters vector $\boldsymbol{\theta}$ in \eqref{eq:model yt} can be estimated without distributional assumptions using ordinary least squares \citep{pinsonVeryShortTermProbabilisticForecasting2012}. With $N+p$ observations, we can express \eqref{eq:model yt} in matrix notation as
\begin{equation} \label{eq:model vec yt}
    \mathbf{y} = \mathbf{Y}_B^\top\boldsymbol{\theta} + \mathbf{z}
    ,\end{equation}
where $\mathbf{y} = [y_{p+1},..., y_{p+N}]^\top$ is the vector observations and $\mathbf{Y}_B$ the predictor matrix with 
\begin{equation}
\mathbf{Y}_B =  \begin{bmatrix}
    1  &  1 & \dots & 1\\
    y_{1}  & y_{2} & \dots & y_{N} \\
    \vdots  & \vdots & \ddots & \vdots \\
    y_{p}  & y_{p+1} & \dots & y_{p+N-1}
    \end{bmatrix} =  \begin{bmatrix}
    \mathbf{y}_{p+1,B_p}  &  \mathbf{y}_{p+2,B_p} & \dots & \mathbf{y}_{N+p,B_p}
    \end{bmatrix} 
    ,\end{equation}
and $\mathbf{z}=[z_{p+1},...,z_{N+p}]^\top$ is the error terms. 
The estimation of $\boldsymbol{\theta}$ by minimising the square error gives the classical result $\boldsymbol{\theta}^\ast_{lse} = (\mathbf{Y}_B\mathbf{Y}_B^\top)^{-1}\mathbf{Y}_B\mathbf{y}$ \citep{Raolinear1973}, which we discuss in Section \ref{sec:RLS}.

The statistical model for wind energy time series introduces additional parameters of interest.
Let random variable $Y_t$ be the current wind power generation at time $t$ awaiting observation. The AR(p) model is then expressed as
\begin{equation}\label{eq:AR p}
    Y_t = \theta_0 + \theta_1 y_{t-1} + \theta_2 y_{t-2} + ... + \theta_p y_{t-p} + Z_t = \mathbf{y}_B^\top\boldsymbol{\theta} + Z_t
    ,\end{equation}
where $Z_t$ the error random variable representing the unexplained variation.
As discussed in part \ref{sec:def}, the generalised logit transformation $Y_t=L_\nu(X_t)$ has the support $[{L_\nu}(\epsilon), {L_\nu}(1-\epsilon)]$.
The error random variable $Z_t$ is assumed following inflated normal distribution $\mathcal{N}^\star (0, \sigma_z^2, {L_\nu}(\epsilon), {L_\nu}(1-\epsilon))$.
Alternatively, we can assume that $Z_t$ follows a normal distribution $\mathcal{N} (0, \sigma_z^2)$ and accumulate all probability density into the boundaries before performing the inverse transformation.
In practice, the error random variables are assumed i.i.d. with shared variance $\sigma_z^2$.

In model \eqref{eq:AR p}, the shape parameter $\nu$ of the generalised logit transformation is assumed fixed throughout the process. 
Estimating this parameter requires more than just the expression based on the transformed observations.
Instead, we must incorporate the relationship defined by the original observed data $\{x_t\}$. 
Re-write \eqref{eq:model yt} as
\begin{equation}
    {L_\nu}(x_t) = \theta_0 + \theta_1 {L_\nu}(x_{t-1}) + \theta_2 {L_\nu}(x_{t-2}) + ... + \theta_p {L_\nu}(x_{t-p}) + z_t,
    \end{equation}
and then \eqref{eq:model vec yt} for observation vector $\mathbf{y}$ becomes
\begin{equation} 
    {L_\nu}(\mathbf{x}) = {L_\nu}(\mathbf{X}_B)^\top\boldsymbol{\theta} + \mathbf{z} 
    .\end{equation}
Note that here, we use ${L_\nu}(\cdot)$ as element-wise transformation to simplify notation.

In summary, we have established the relationship between the current value of the transformed data with their past values. 
Our primary focus is on estimating the model parameters $\boldsymbol{\theta}$, the variance of observation error $\sigma_z^2$, and the shape parameter $\nu$.

\section{Classical Estimation Methods}\label{sec:background}

In this section, we review two parameter estimation methods for the AR(p) model introduced in Section~\ref{sec:model}: the recursive least squares (RLS) and the iterative Newton-Raphson approaches.
The RLS method provides an adaptive update for the model parameters $\boldsymbol{\theta}_t$, while keeping the shape parameter fixed.
To enable adaptive estimation of the shape parameter, \cite{Amandine2021} proposed a method based on the Newton-Raphson optimisation step applied to the negative log-likelihood function.

\subsection{Recursive Least Squares (RLS) Method}\label{sec:RLS}

Consider the estimation of $\boldsymbol{\theta}_t$ for an AR(p) model at time $t=N+p$.
The least squares method looks for the optimal $\boldsymbol{\theta}^\ast_t$ by minimising the sum of the squares of the fitting error
\begin{equation}
    \boldsymbol{\theta}^\ast_{t,lse} = \arg\min_{\boldsymbol{\theta}} \sum_{i=1}^N z_{p+i}^2 
    =  \arg\min_{\boldsymbol{\theta}} \sum_{i=1}^N (y_{p+i} - \mathbf{y}_{p+i,B}^\top \boldsymbol{\theta})^2 
    =  \arg\min_{\boldsymbol{\theta}} \lVert \mathbf{y}_t - \mathbf{Y}_{t,B}^\top\boldsymbol{\theta} \rVert^2_2 
    \end{equation}
where $\mathbf{y}_{p+i,B} = [1, y_{p+i-1},...,y_{p+i-p}]^\top$ contains the $p$ back-shifted past observation values with intercept, and $\mathbf{Y}_{t,B}$ is equivalent to $\mathbf{Y}_{B}$ containing data available at time $t$.
It can be shown that
\begin{equation}
\label{eq:RLS solution} 
    \boldsymbol{\theta}^\ast_{t,lse} = (\mathbf{Y}_{t,B}^{}\mathbf{Y}_{t,B}^\top)^{-1}\mathbf{Y}_{t,B}^{}\mathbf{y}_t
    .\end{equation}

The recursive version involves updating the model each time a new forecast is made (i.e., when a new observation becomes available). 
The key idea behind recursive least squares is to avoid recomputing the full linear equation
$\mathbf{Y}_{t,B}^{}\mathbf{Y}_{t,B}^\top\boldsymbol{\theta}^\ast_{t,lse} = \mathbf{Y}_{t,B}^{}\mathbf{y}_t$ at each refitting step.
Assume that at times $t$ and $t+M$ there are $N+p$ and $N+p+M$ observations available, respectively.
The least square estimation for $\boldsymbol{\theta}^\ast_{t}$ and $\boldsymbol{\theta}^\ast_{t+M}$ are
\begin{align}
\label{eq:raw rls update} 
    \boldsymbol{\theta}^\ast_{t,lse} &= (\mathbf{Y}_{t,B}^{}\mathbf{Y}_{t,B}^\top)^{-1}\mathbf{Y}_{t,B}^{}\mathbf{y}_t \\
\label{eq:raw rls update 2} 
    \boldsymbol{\theta}^\ast_{t+M,lse} &= (\mathbf{Y}_{t+M,B}^{}\mathbf{Y}_{t+M,B}^\top)^{-1}\mathbf{Y}_{t+M,B}^{}\mathbf{y}_{t+M}
    .\end{align}
Notice that
\begin{align}
\label{eq:rls P_t+M resursive} 
    \mathbf{Y}_{t+M,B}^{}\mathbf{Y}_{t+M,B}^\top &= \mathbf{Y}_{t,B}^{}\mathbf{Y}_{t,B}^\top + \mathbf{\tilde{Y}}_{t+M,B}^{}\mathbf{\tilde{Y}}_{t+M,B}^\top 
    \\
\label{eq:rls Y_t+M y decomposition} 
    \mathbf{Y}_{t+M,B}^{}\mathbf{y}_{t+M} &=  \mathbf{Y}_{t,B}^{}\mathbf{y}_{t} + \mathbf{\tilde{Y}}_{t,B}^{}\mathbf{\tilde{y}}_{t}
    \end{align}
where $\mathbf{\tilde{Y}}_{t+M,B}^{} = [\mathbf{y}_{t+1,B}...,\mathbf{y}_{t+M,B}]$ and $\mathbf{\tilde{y}}_{t} = [y_{t+1},...,y_{t+M}]$ are newly obtained observations.
Let
\begin{equation}
    \mathbf{P}_t=\mathbf{Y}_{t,B}^{}\mathbf{Y}_{t,B}^\top
    ,\quad
    \mathbf{P}_{t+M}=\mathbf{Y}_{{t+M},B}^{}\mathbf{Y}_{{t+M},B}^\top
    ,\quad
    \boldsymbol{\tilde{\epsilon}}_t= \mathbf{\tilde{y}}_{t+M}^{} - \mathbf{\Tilde{Y}}_{t+M,B}^{\top}\boldsymbol{\theta}^\ast_{t,lse}
    .\end{equation}
Using the approach from \cite{Ljung} and the Sherman–Morrison–Woodbury formulas \citep{max1950inverting}, 
the optimal estimation at time $t+M$, $\boldsymbol{\theta}^\ast_{t+M,lse}$, can then be expressed as
\begin{align}
\label{eq:rls theta recursive final 1}
    \boldsymbol{\theta}^\ast_{t+M,lse} &= \boldsymbol{\theta}^\ast_{t,lse} + \mathbf{P}_{t+M}^{-1}\mathbf{\tilde{Y}}_{t+M,B}\boldsymbol{\tilde{\epsilon}}_t = \boldsymbol{\theta}^\ast_{t,lse} + \mathbf{P}_t^{-1}
    \mathbf{\tilde{Y}}_{t+M,B}^{}
    (I_M + \mathbf{\tilde{Y}}_{t+M,B}^{\top} \mathbf{P}_t^{-1}\mathbf{\tilde{Y}}_{t+M,B}^{})^{-1}
    {\tilde{\epsilon}}_t
    \end{align}
where $I_M$ is the identity matrix with M dimensions.

To initialise $\mathbf{P}$, we could consider any positive definite matrix; for instance, an identity matrix.
However, the choice of the initial $\mathbf{P}$ influences the recursive update behaviour, which we discuss in section \ref{sec:Sensitivity Analysis}.
In contrast, the initialisation of $\boldsymbol{\theta}$ has a lesser impact on recursive learning.
In our case, a zero vector is sufficient for the model to converge to the optimal value.

A more stable recursive update approach involves using a forgetting factor, denoted as $\lambda \in (0,1)$, which is typically chosen close to $1$.
This is because, as data accumulate, the matrix $\mathbf{P}$ tends to have large eigenvalues, which can lead to numerical instabilities.
Additionally, without a forgetting factor, all past data contribute equally to parameter updates, which contradicts the general expectation that more recent data should have a higher impact on model parameter adaptation.
Introducing a forgetting factor modifies the matrix update of $\mathbf{P}$ as follows:
\begin{equation} \label{eq: P smoothing}
    \mathbf{P}_{t+M} = \lambda^M \mathbf{P}_t +\mathbf{\tilde{Y}}_{t+M,B}^{}\mathbf{\tilde{Y}}_{t+M,B}^\top
    .\end{equation}
When using a forgetting factor, the update result may differ between a one-by-one update and an M-by-M step update, depending on how data are received and fed into the model.
Thus, the choice of update batch size $M$ should be guided by the specific application.
The recursive equation \eqref{eq: P smoothing} can be interpreted as a smoothing process \citep{Hyndman2008}, and the choice of the forgetting factor can be related to the number of effective observations, denoted $n_\lambda$, via $ \lambda = 1 - 2/({n_\lambda + 1})$.
The value of $\lambda$ mainly depends on the matrix $\mathbf{\tilde{Y}}_{t+M,B}^{}$.
Choosing $\lambda$ too small or too large can lead to numerical issues due to ill-conditioned matrix $\mathbf{P}_{t+M}$, either with excessively large or overly small eigenvalues.
With the above recursive update equation, model parameter estimation can now be performed in an adaptive fashion.

The classical LS variance estimator is
\begin{equation}
    \hat{\sigma}_z^2 = \frac{1}{M} \sum_{i=1}^M (y_{p+i} - \mathbf{y}_{p+i,B}^\top \boldsymbol{\theta}^\ast_{lse})^2 
    .\end{equation}
\cite{pinsonVeryShortTermProbabilisticForecasting2012} also provides a method to update this estimate adaptively.
The key idea is that for each deterministic forecast $\hat{y}_t \in [0,1]$, a weight $w^\ast_t \in [0,1]$ is computed as $w^\ast_t = 1- (1-\lambda)\cdot4\hat{y}_t(1-\hat{y}_t)$.
The estimator is then updated as
\begin{equation}
    \hat{\sigma}_{z,t+1}^2 = (1-w^\ast_t)\hat{\sigma}_{z,t}^2 + w^\ast_t(y_{t+1} - \hat{y}_{t+1})^2
    .\end{equation}
The calculation of $w^\ast_t$ depends on the forgetting factor $\lambda$,
meaning that the choice of $\lambda$ also influences the variance estimation update.
However, electing an appropriate weighting scheme should be determined on a case-by-case basis.

\subsection{Iterative Newton-Raphson Method}\label{sec:NR}

Consider model parameters $\boldsymbol{\theta}_t$, the variance $\sigma^2_{z,t}$ and the shape parameter $\nu_t$ at time $t$, estimated from the dataset $\mathbf{x}_t$ and $\mathbf{X}_{t,B}$.
After receiving $M$ new observations $\tilde{\mathbf{x}}_{t+M}$ and constructing $\tilde{\mathbf{X}}_{t+M,B}$, the likelihood of the full dataset can be expressed as
\begin{equation}
p({{\mathbf{x}}_{t+M}}|{\mathbf{X}}_{t+M,B},\boldsymbol{\theta}_t,\sigma^2_{z,t},\nu_t) = 
    {(2\pi \sigma^2_{z,t})}^{-\frac{t+M-p}{2}}
    \prod_{i=p+1}^{t+M} \frac{\nu_t}{{x}_{i}(1-{x}_{i}^{\nu_t})} \\
    \exp \Bigl\{ -\frac{\lVert {L}_{\nu_t}({\mathbf{x}_{t+M}}) - {L}_{\nu_t}({\mathbf{X}}_{t+M,B})^\top \boldsymbol{\theta}_t \rVert^2}{2\sigma^2_{z,t}} \Bigl\}
    .\end{equation}
The negative log-likelihood, modified with exponential forgetting for the full set of observation is then \citep{Amandine2021}
\begin{align}
    l_{t+M}(\boldsymbol{\theta}_t,\sigma^2_{z,t},\nu_t) &= -(1-\lambda) \sum_{i=p+1}^{t+M}\lambda^{t+M-i}\ln p({{{x}}_{i}}|{\mathbf{X}}_{i,B},\boldsymbol{\theta}_t,\sigma^2_{z,t},\nu_t)
    .\end{align}
Here, the terms $\lambda^{t+M-p}, \dots, \lambda^0$, serve as exponential forgetting weights for the observations, where $\lambda$ is the forgetting factor, consistent with its role in the RLS method.

The Newton-Raphson step is applied to all parameters. 
Let $w_t = [\boldsymbol{\theta}_t^\top,\sigma^2_{z,t},\nu_t]^\top$. 
The one-step update from $w_t$ to $w_{t+1}$, i.e., setting $M=1$, is given by 
\citep{Amandine2021} as
\begin{equation}
    w_{t+1} = w_t - {\nabla^2_{w} l_{t+1}(w_t)}^{-1}{\nabla_{w} l_{t+1}(w_t)}
    .\end{equation}
An approximation for the quadratic term $\nabla^2_{w}l_{t+1}(w_t)$ is provided by \cite{Amandine2021}:
\begin{equation}\label{eq: NR Update step}
    \nabla^2_{w} l_{t+1}(w_t) = \lambda \nabla^2_{w}l_t(w_t) + (1-\lambda)\boldsymbol{h}^{}_{t+1} \boldsymbol{h}_{t+1}^\top
    ,\end{equation}
where $\boldsymbol{h}^{}_{t+1} = \nabla_{w}\ln p({{{x}}_{t+1}}|{\mathbf{X}}_{t+1,B},w_t)$.

\section{Proposed modelling framework}\label{sec:ourmodel}

As stated in Section~\ref{sec:intro}, the Bayesian framework naturally lends itself to probabilistic forecasting, enabling us to quantify uncertainty in both the response variable and the model parameters. 
We here introduce adaptive Bayesian updates for our model parameters assuming fixed and varying $\nu$.
To keep computational costs manageable, our adaptive update with varying $\nu$
integrates Bayesian estimation of $\boldsymbol{\theta}$ with point-wise updates for the shape parameter $\nu$ based on Maximum A Posteriori (MAP) estimation.
In this way, the parameters $\boldsymbol{\theta}$, $\sigma^2$ and $\nu$ are updated alternately, which is analogous to the coordinate descent method \citep{MESSNER20191485}.

\subsection{Adaptive Bayesian Method with Fixed Shape Parameter}\label{sec:Bayes}

Consider the newly received transformed data vector $\mathbf{\tilde{y}}_{t+M}$ at time $t+M.$
Assuming a Gaussian prior distribution on $\boldsymbol{\theta}$ and a Gamma distribution on the inverse of the model variance, we obtain the following hierarchical structure:
\begin{align}\label{eq:hier_bayes_fixednu}
    p(\mathbf{\tilde{y}}_{t+M}|\mathbf{\tilde{Y}}_{t+M,B},\boldsymbol{\theta}_t,\sigma_{z,t}) &= \prod_{i=t+1}^{t+M} p(y_i|\mathbf{y}_{i,B},\boldsymbol{\theta}_t,\sigma_{z,t}), \nonumber\\
    \boldsymbol{\theta}=[\theta_0,...,\theta_p]^\top &\sim \mathcal{N}(\boldsymbol{\mu}_{\boldsymbol{\theta}}, \Sigma_{\boldsymbol{\theta}}),\nonumber\\
    \sigma_{z,t}^{-2} &\sim \mathcal{G}(\alpha, \beta),
    \end{align}
where $p(\mathbf{\tilde{y}}_{t+M}|\mathbf{\tilde{Y}}_{t+M,B},\boldsymbol{\theta}_t,\sigma_{z,t})$ denotes the likelihood for the vector of observations $\mathbf{\tilde{y}}_{t+M} = [y_{t+1},...,y_{t+M}]$ and $p(y_i|\mathbf{y}_{i,B},\boldsymbol{\theta}_t,\sigma_{z,t})$ denotes the conditional distribution of $y_i$ given $\mathbf{y}_{i,B}$, which can be a Gaussian or inflated Gaussian distribution (as discussed in Section~\ref{sec:def}) with mean $\boldsymbol{\theta}_t$ and standard deviation $\sigma_{z,t}$, and $\boldsymbol{\mu}_{\boldsymbol{\theta}}, \Sigma_{\boldsymbol{\theta}}, \alpha$ and $\beta$ are fixed hyperparameters which application-specific values; see Section~\ref{sec:exp}.
Note that the likelihood in~\eqref{eq:hier_bayes_fixednu} can be seen as the probabilistic forecast for time $t+M$.
The conjugated model in~\eqref{eq:hier_bayes_fixednu} assumes that observations are conditionally independent given the parameters, and that the priors for $\boldsymbol{\theta}$ and $\sigma_{z,t}^{-2}$ are independent.

The Cumulative Distribution Function (CDF) in the transformed domain is transformed back to its original domain $[\epsilon, 1 - \epsilon]$ to obtain the final probabilistic forecasts in the form of the predictive CDF.
Using CDF for probabilistic forecasting is more convenient than the PDF here as handling  the mixture of probability density and probability masses in the CDF is much simpler computationally.

Assume now that at time $t$, $M$ new observations have been received.
Let $P_{\boldsymbol{\theta}}= \Sigma_{\boldsymbol{\theta}}^{-1}$ and $\mathbf{P}_Z= \diag(\sigma_{z,0}^{-2},...,\sigma_{z,0}^{-2})_M$ be an M-by-M square matrix.
Then, posterior updates for the model parameters can be expressed as
\begin{align}
\label{bayes: precision update} 
    \mathbf{P}_{\boldsymbol{\theta},t}^{\ast} &= \mathbf{P}_{\theta,t}^{} + \mathbf{\tilde{Y}}_{t,B}^{}\mathbf{P}_{Z,t}\mathbf{\tilde{Y}}_{t,B}^{} \\
\label{bayes: original mu update}  
    \boldsymbol{\mu}_{\boldsymbol{\theta},t}^{\ast} &= \mathbf{P}_{\boldsymbol{\theta},t}^{\ast-1}(\mathbf{\tilde{Y}}_{t,B}^{}\mathbf{P}_{Z,t} \mathbf{\tilde{y}}_t+ \mathbf{P}_{\theta,t}^{} \boldsymbol{\mu}_{\boldsymbol{\theta},t}) \\
\label{bayes: alpha update} %
    \alpha_{t}^{\ast} &= {\alpha_t} + \frac{M}{2} \\
\label{bayes: beta update} %
    \beta_{t}^{\ast} &= {\beta_t}
    + \frac{1}{2}(\mathbf{\tilde{y}}_{t}^\top \mathbf{\tilde{y}}_{t}^{}
    + \boldsymbol{\mu}_{\boldsymbol{\theta},t}^\top  \mathbf{P}_{\boldsymbol{\theta},t,K}^{} \boldsymbol{\mu}_{\boldsymbol{\theta},t}
    -\boldsymbol{\mu}_{\boldsymbol{\theta},t}^{\ast\top}  \mathbf{P}_{\boldsymbol{\theta},t,K}^{\ast} \boldsymbol{\mu}_{\boldsymbol{\theta},t}^{\ast})
    .\end{align}
Here,
$\mathbf{P}_{\boldsymbol{\theta},t,K}^{}= \mathbf{P}_{\boldsymbol{\theta},t}^{}\mathbf{P}_{Z,t}^{-1}$ and $\mathbf{P}_{\boldsymbol{\theta},t,K}^{\ast}= \mathbf{P}_{\boldsymbol{\theta},t}^{\ast}\mathbf{P}_{Z,t}^{-1}$.

To enable adaptive estimation, we propose an alternating update between the prior and posterior at consecutive time steps. 
Let the prior and posterior distributions at time $t$ and $t+M$ be 
\begin{align}
    \boldsymbol{\theta}_t &\sim \mathcal{N}(\boldsymbol{\mu}_{\boldsymbol{\theta},t}, \Sigma_{\boldsymbol{\theta},t}) ,\quad P_{z,t} \sim \mathcal{G}(\alpha_t, \beta_t) ,\\
    \boldsymbol{\theta}_t^\ast &\sim \mathcal{N}(\boldsymbol{\mu}_{\boldsymbol{\theta},t}^\ast, \Sigma_{\boldsymbol{\theta},t}^\ast), \quad P_{z,t}^\ast \sim \mathcal{G}(\alpha_t^\ast, \beta_t^\ast),
    \\
      \boldsymbol{\theta}_{t+M} &\sim \mathcal{N}(\boldsymbol{\mu}_{\boldsymbol{\theta},{t+M}}, \Sigma_{\boldsymbol{\theta},{t+M}}) ,\quad P_{z,{t+M}} \sim \mathcal{G}(\alpha_{t+M}, \beta_{t+M}) ,\\
    \boldsymbol{\theta}_{t+M}^\ast &\sim \mathcal{N}(\boldsymbol{\mu}_{\boldsymbol{\theta},{t+M}}^\ast, \Sigma_{\boldsymbol{\theta},{t+M}}^\ast), \quad P_{z,{t+M}}^\ast \sim \mathcal{G}(\alpha_{t+M}^\ast, \beta_{t+M}^\ast)
    .\end{align}
A naive adaptive approach is to set $\boldsymbol{\theta}_{t+M} = \boldsymbol{\theta}_t^\ast$ and $P_{z,{t+M}} = P_{z,t}^\ast$, meaning the prior distribution parameters at the current time $t+M$ are taken as the posterior distribution parameters from the previous time $t$.
This ensures the model is continuously updated with newly received data while retaining a reference to all historical data, including the initial prior.

However, the eigenvalue explosion issue discussed in Section \ref{sec:RLS} must also be addressed here.
Rather than directly reusing the posterior distribution, we introduce decay (or forgetting) factors $\lambda_{\boldsymbol{\theta}} \in (0,1)$ and $\lambda_z \in (0,1)$ to adjust the prior distribution at time $t+M$ based on the posterior distribution at time $t$:
\begin{align}
    \boldsymbol{\theta}_{t+M} &\sim \mathcal{N}(\boldsymbol{\mu}_{\boldsymbol{\theta},{t+M}}, \Sigma_{\boldsymbol{\theta},{t+M}}) \equiv \mathcal{N}(\boldsymbol{\mu}_{\boldsymbol{\theta},{t}}, \lambda_{\boldsymbol{\theta}}^{-M}\Sigma_{\boldsymbol{\theta},{t}}) \\
    P_{z,{t+M}} &\sim \mathcal{G}(\alpha_{t+M}, \beta_{t+M}) \equiv \mathcal{G}(\lambda_z\alpha_{t}, \lambda_z\beta_{t})
    .\end{align}
The choice of decay factor values also depends on the characteristics of the data.

Notice that in the updated equations \eqref{bayes: precision update} and \eqref{bayes: original mu update}, the term $\mathbf{P}_{\boldsymbol{\theta},t}^{-1}$ involves the inversion of the sum of two positive definite matrices, aligning with equations \eqref{eq:rls P_t+M resursive} and \eqref{eq:rls Y_t+M y decomposition}.
Applying the Sherman–Morrison–Woodbury formulas again, the update for $\boldsymbol{\mu}_{\boldsymbol{\theta},t}^{\ast}$ in \eqref{bayes: original mu update} can be rewritten as
\begin{equation}
\label{bayes: mu update with error version}
    \boldsymbol{\mu}_{\boldsymbol{\theta},t}^{\ast} = \boldsymbol{\mu}_{\boldsymbol{\theta},t} +  \mathbf{P}_{\boldsymbol{\theta},t}^{-1}\mathbf{\tilde{Y}}_{t,B}^{}
    (\mathbf{P}_{Z,t}^{-1} + \mathbf{\tilde{Y}}_{t,B}^{\top}\mathbf{P}_{\boldsymbol{\theta},t}^{-1}\mathbf{\tilde{Y}}_{t,B}^{})^{-1}
    (\mathbf{\tilde{y}}_t - \mathbf{\tilde{Y}}_{t,B}^{\top}\boldsymbol{\mu}_{\boldsymbol{\theta},t})
    .\end{equation}
For time $t+M$, this formulation directly coincides with \eqref{eq:rls theta recursive final 1} when setting $\mathbf{P}_{Z,t}^{-1} = \mathcal{I}_M$.
This reveals the intrinsic connection between the RLS approach and the Bayesian framework proposed here.
Moreover, this update also overlaps with the Kalman filter method \citep{Grewal2014}, where $\boldsymbol{\mu}$ corresponds to the state variables under estimation.
One key advantage of our proposed Bayesian approach is that, by adaptively estimating $\mathbf{P}_{Z,t}$, the incorporation of new data is adjusted with respect to $\mathbf{P}_{Z,t}$, as reflected in equations \eqref{bayes: precision update} and \eqref{bayes: original mu update}.

\subsection{Adaptive Bayesian Method with Varying Shape Parameter}\label{sec:AdaBayes}

Consider the newly received data $\tilde{\mathbf{x}}_{t+M}$ at time $t+M$.
With the posterior parameter estimates $\boldsymbol{\theta}^{\ast}_{t+M}$ and $P^{\ast,-1}_{z,t+M} = \sigma^{\ast,2}_{z,t+M}$, the posterior likelihood for the new data can be expressed as
\begin{multline}
p(\tilde{\mathbf{x}}_{t+M}|\tilde{\mathbf{X}}_{t+M,B},\boldsymbol{\theta}^{\ast}_{t+M},\sigma^{\ast,2}_{z,t+M},\nu_{t+M}) = 
    {(2\pi \sigma^{\ast,-2}_{z,t+M})}^{-\frac{M}{2}}
    \prod_{i=t+1}^{t+M} \frac{\nu_{t+M}}{{x}_{i}(1-{x}_{i}^{\nu_{t+M}})} \\ 
    \cdot \exp \Bigl\{ -\frac{\lVert {L}_{\nu_{t+M}}({\tilde{\mathbf{x}}_{t+M}}) - {L}_{\nu_{t+M}}({\tilde{\mathbf{X}}_{t+M,B}})^\top \boldsymbol{\theta}^{\ast}_{t+M} \rVert^2}{2\sigma^{\ast,2}_{z,t+M}} \Bigl\}
    .\end{multline}
The negative log-likelihood $\tilde{L}(\nu) := -\ln p(\tilde{\mathbf{x}}_{t+M}|\tilde{\mathbf{X}}_{t+M,B},\boldsymbol{\theta}^{\ast}_{t+M},\sigma^{\ast,2}_{z,t+M},\nu)$ is then used to construct our objective function to update $\nu_{t+M}$.

Let $P^{\ast}_{z,t+M}$ be the updated precision for the model parameter $\boldsymbol{\theta}^{\ast}_{t+M}$.
By definition, $P^{\ast}_{z,t+M}$ is strictly positive definite, so we can apply the Cholesky decomposition to $P^{\ast}_{z,t+M}$:
\begin{equation}
\label{eq:Cholesky}
    P^{\ast}_{z,t+M} = k^2\mathbf{L}\mathbf{L}^\top
    ,\end{equation}
where $\mathbf{L}$ is a lower triangular matrix with real and positive diagonal entries, and $k$ is a positive regularisation parameter.
By setting $\mathbf{y}^\ast = \mathbf{L}^\top \boldsymbol{\mu}_{\boldsymbol{\theta},{t+M}}^\ast$, the least squares estimation of the response data $\mathbf{y}^\ast$ and predictors $\mathbf{L}$ is $\boldsymbol{\theta}^{\ast}_{t+M}=\boldsymbol{\mu}_{\boldsymbol{\theta},{t+M}}^\ast$.
From equation \eqref{eq:RLS solution}, we know that $\boldsymbol{\theta}^{\ast}_{t+M} = (\mathbf{L}\mathbf{L}^\top)^{-1}\mathbf{L}\mathbf{y}^\ast=(\mathbf{L}\mathbf{L}^\top)^{-1}\mathbf{L}\mathbf{L}^\top \boldsymbol{\mu}_{\boldsymbol{\theta},{t+M}}^\ast=\boldsymbol{\mu}_{\boldsymbol{\theta},{t+M}}^\ast$.
We define the terms
\begin{align}
    \begin{cases} 
        \mathbf{x}^\ast:= {L}_{\nu_{t+M}}^{-1}(\mathbf{y}^\ast), & \mathbf{X}^\ast_{B}:= {L}_{\nu_{t+M}}^{-1}(\mathbf{L})\\
        \mathbf{y}_{\nu}:= {L}_{\nu}(\mathbf{x}^\ast), & \mathbf{Y}_{B,\nu}:= {L}_{\nu}(\mathbf{X}^\ast_{B})\\
    \end{cases},
\end{align}
which represent the reconstructed data from $\boldsymbol{\theta}^{\ast}_{t+M}$ in both the original and transformed domains.
Then, $\nu = \nu_{t+M}$ maximises the likelihood of the transformed reconstructed data
\begin{equation}
    p(\mathbf{y}_{\nu}|\mathbf{Y}_{B,\nu},\boldsymbol{\theta}^{\ast}_{t+M},\sigma^{\ast,2}_{z,t+M}) = 
    {(2\pi \sigma^{\ast,-2}_{z,t+M})}^{-\frac{p}{2}}
    \exp \Bigl\{ -\frac{\lVert \mathbf{y}_{\nu} - \mathbf{Y}_{B,\nu}^\top \boldsymbol{\theta}^{\ast}_{t+M} \rVert^2}{2\sigma^{\ast,2}_{z,t+M}} \Bigl\}
    .\end{equation}  
With the negative log likelihood ${L}^\ast(\nu) := -\ln p(\mathbf{y}_{\nu}|\mathbf{Y}_{B,\nu},\boldsymbol{\theta}^{\ast}_{t+M},\sigma^{\ast,2}_{z,t+M})$, the updated $\nu_{t+M}$ can be found by
\begin{align}
\label{eq:bayes nu} 
    \hat{\nu}_{t+M} &= \arg\min_{\nu} \Bigl\{{L}^\ast(\nu) + \tilde{L}(\nu) \Bigl\} ,\\
    {\nu}_{t+M}^\ast &:= (1-\gamma){\nu}_{t+M} + \gamma \hat{\nu}_{t+M}
    ,\end{align} 
with the learning rate $\gamma \in (0,1)$.
Equation \eqref{eq:bayes nu} consists of two parts: one from the newly received data and the other from the reconstructed data.
$\tilde{L}(\nu)$ is the sum of $M$ negative log-likelihood terms, whilst ${L}^\ast(\nu)$ consists of the sum of $p$ terms.
During the initial inference or when inference is performed on a substantial historical dataset, $M$ could be large compared to $p$.
However, during the adaptive forecasting and update phase, $M$ could be of similar magnitude to $p$.
The relative sizes of $M$ and $p$ reflect the importance of the received data and the reconstructed data.
For large $M$, $\tilde{L}(\nu)$ will dominate the objective function and vice versa.

\section{Case Study}\label{sec:exp}

In this section, we will assess the methods discussed using wind power generation data collected from wind farms in the UK.
Sections \ref{sec:DDP} and \ref{sec: DQA} will provide a brief overview of the data and evaluate its quality. After detailing the model and method implementation in Section \ref{sec:MIIF}, we will present and discuss various evaluations, including metrics such as the Continuous Ranked Probability Score (CRPS) and Skill Score, along with reliability and sensitivity analyses.

\subsection{Data Description and Preprocessing}\label{sec:DDP}

We evaluated the proposed methods using data from 128 wind farms available from the Balancing Mechanism Reporting Service (BMRS) API provided by Elexon\textsuperscript{\textregistered} (replaced by Insight Solution from May 2024).
The data cover electricity production between years 2020 to 2023, with time resolution of 30 minutes.
For each wind farm, data comprise the average power production (in MW), the half-hour energy instructions from the system operator (in MWh) to reduce (Bid Acceptance Volume) or increase (Offer Acceptance Volume) production.
Power production is corrected using BAV and OAV in order to remove the impact of system operator interventions and maximise the volume of data available.

We train and evaluate models in a rolling fashion using data from one year to initially train each model, and data from the following year to evaluate the forecasting results, including adaptive updates where relevant.
Thus, except for wind farms which lack data for certain years due to them being (de-)commissioned, each wind farm has three train-test scenarios, which are (2020, 2021), (2021, 2022) and (2022, 2023).
For each train-test scenario and wind farm, data were re-scaled into $[0,1]$ according to the maximum wind power generation of the training data after removing outliers.
$\epsilon$ is set to 0.005 to constrain the data into $[\epsilon, 1-\epsilon]$, as described in Section \ref{sec:def}.
Finally, the data is reshaped from a time series format into observation-predictor pairs.
If any values are missing, then any corresponding observation-predictor pairs will not be used in model training nor updates, and the corresponding forecasts will not be produced or evaluated. Missing data is discussed further in the next section.

\subsection{Data Quality Analysis}\label{sec: DQA}

\begin{figure}
    \centering
    \includegraphics[width=1\linewidth]{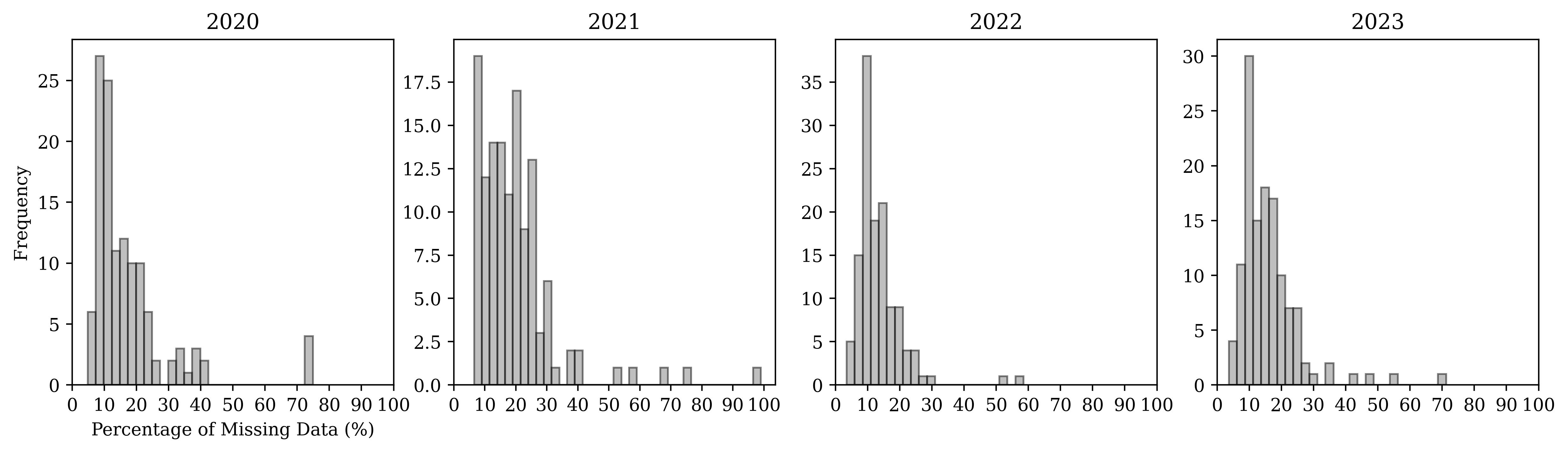}
    \caption{Histogram of percentage of missing data over the four years. }
    \label{fig:nan des}
    \end{figure}

All wind farms have some missing data, the extent of which is shown in Figure \ref{fig:nan des} by year.
On average, wind farm power generation data contains approximately $10\%$ missing data.
For certain wind farms, over $30\%$ of data is missing, which could be attributed to two main reasons:
the wind farm was newly commissioned in that year; or,
extended periods of no production due to outages of the wind farm itself or essential electrical equipment.
When wind farms are commissioned, their installed capacity may appear to grow over an extended period of time, especially for large offshore wind farms. This makes rescaling data challenging and can lead to spurious forecasting results.

\begin{figure}
    \centering
    \includegraphics[width=1\linewidth]{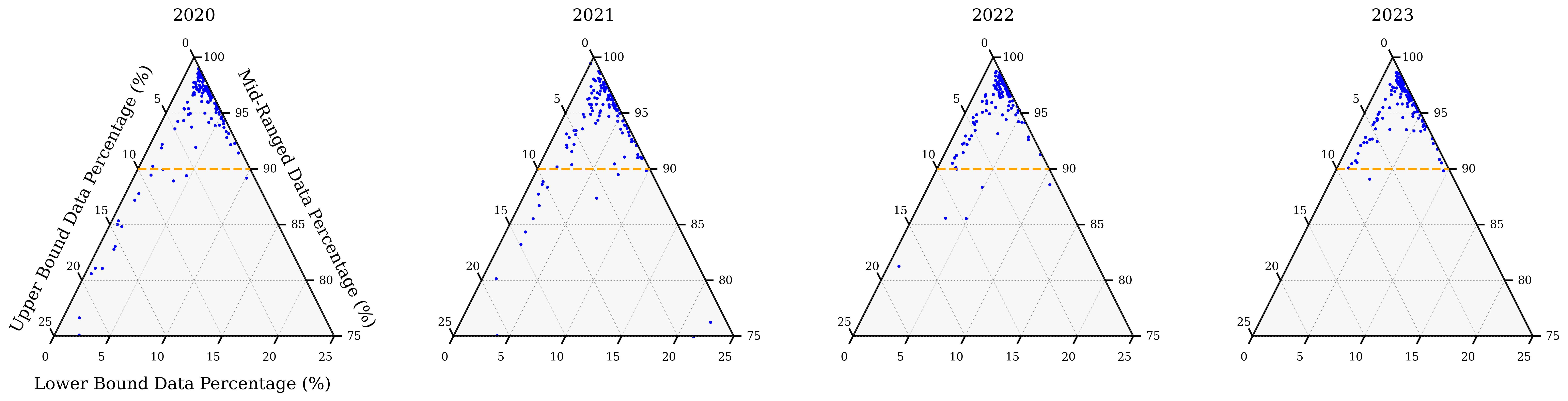}
    \caption{Ternary chart \citep{pythonternary} for lower and upper bound data over four years, excluding missing data.}
    \label{fig:bound des}
    \end{figure}

We examine the proportion of data on the boundaries for each wind farm.
This part of the analysis was conducted after excluding missing data.
Data on the lower bound $\epsilon$ and upper bound $1-\epsilon$ are modelled by a probability mass, whether conditional on the predictor or not.
Additionally, the presence of the bound value affects the suitability of the Gaussian distribution assumption on transformed observation data.
Figure \ref{fig:bound des} illustrates the proportion of bound data by wind farm and year.
The bottom, left and right axes indicate the proportion of data equal to the lower bound $\epsilon$, within the interval $(\epsilon, 1-\epsilon)$, and equal to the upper bound $1-\epsilon$, respectively.
In summary, all the wind farms we have studied contain at least 2\% lower bound data. 
Most wind farms contain more than $90\%$ of their data between bounds.
For the upper bound proportion, the majority of wind farms contain no more than $2\%$.
Certain wind farms have between 5\% and 10\% upper bound data.
Wind farms may operate under conditions or restrictions that cause them to function at levels below the observed yearly maximum power, which are then classified as mid-range data.

\begin{figure}
    \centering
    \includegraphics[width=1\linewidth]{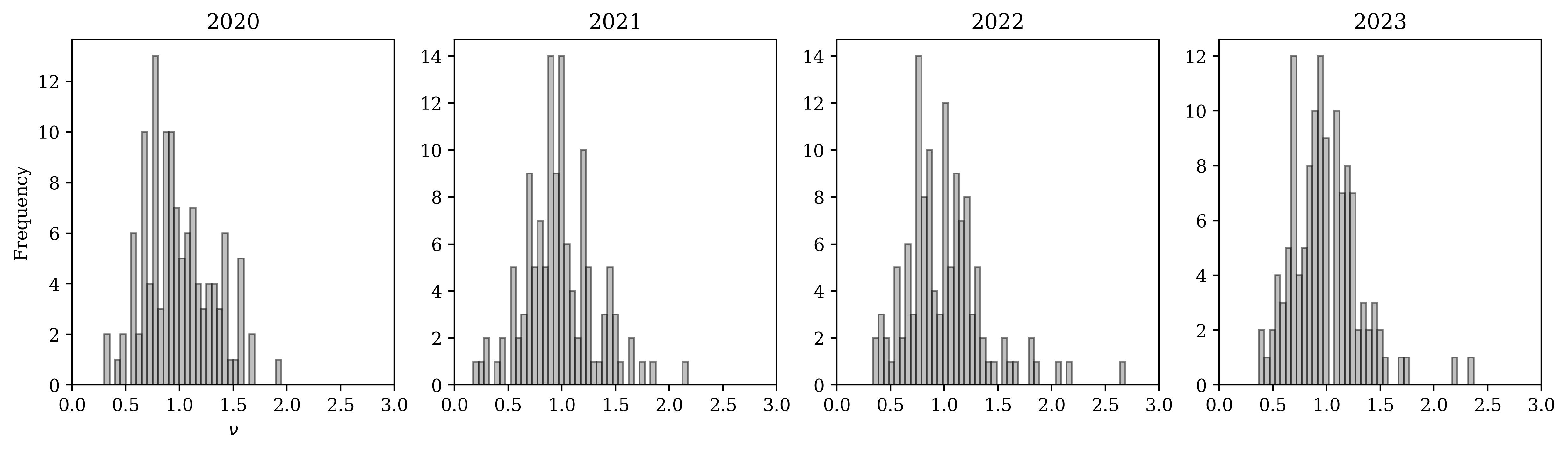}
    \caption{Histogram of estimated shape parameter across 4 years. The x-axis is the bin of the estimated shape parameter, while the y-axis shows the frequency of wind farms corresponding to each bin.}
    \label{fig:nu des}
    \end{figure}

Finally, we performed a brief investigation into the shape parameter distribution across all wind farms, as previous works by, e.g., \cite{pinsonVeryShortTermProbabilisticForecasting2012} and \cite{Amandine2021} only report results for individual wind farms.
The shape parameter $\nu$ is estimated for each year and  wind farm, using the AR-$L_\nu$ method (see Section \ref{sec:MIIF}).
Figure \ref{fig:nu des}  demonstrates variation of shape parameter across wind farms and over the years.
Although $\nu$ is estimated as a single value for the entire year in this analysis, the results suggest that, for a large proportion of wind farms, $\nu=1$ is a reasonable initialisation.
Another potential implication is that, if the $\nu$ does not vary a lot in reality, then the methods that adaptively update $\nu$ may not significantly affect forecast performance.

\subsection{Implementation, Initialisation and Forecasting}\label{sec:MIIF}

According to the results of the data quality analysis, there are outlier wind farms for which the pre-processing is ineffective due to the unusual variation in apparent wind farm capacity. In practice, forecasters would likely have information regarding wind farm availability that is not available to us here; therefore, those wind farms will be removed from the case study.
For each wind farm $w$, we used the variance of the rolling max with a window size of three months $\sigma^2_w$ and the indicator calculated via quantiles $w_q := 3 \times (q_{0.95}-q_{0.05})$ to identify outlier wind farms.
We only keep the wind farms that satisfy the criteria $w_q>\sigma^2_w$.
In total, 27 outlier wind farms were removed, leaving 101 wind farms in our case study. Results for the full dataset are included in Appendix \ref{apd:CRPS}.

Seven methods have been implemented to evaluate and compare against the proposed method.
All except for the persistence model are autoregressive-type (AR) models. The order (lags) are selected according to the partial autocorrelation function (PACF, \cite{Shumway2017}). A minimal order (lag) is set to $p=1$, and maximal order (lag) is set to $p=6$.
Naming and summaries of each approach are listed below.

\begin{description}

\item[Persistence]
The persistence method serves as a naive benchmark. It is a standard benchmark that performs remarkably well for very short-term wind power forecasting \citep{tawnReviewVeryshortterm2022}.
The sample variance $\sigma_z^2$ of forecast residuals will be used to parametrise an inflated Gaussian distribution for probabilistic forecasts .
The persistence method provides probabilistic forecasts directly in the original (untransformed) domain.
All other methods will provide probabilistic forecasts on the transformed domain and will be transformed back to the original domain for evaluation.

\item[AR-$L$] AR with standard logit data transformation, i.e., $\nu=1$.
AR model will be fitted on the transformed data using ordinary least squares. Error variance $\sigma_z^2$ is estimated based on residuals.

\item[AR-$L_\nu$] AR with generalised logit data transformation.
Model fitted on transformed data. Coordinate descant optimisation is used to estimate model parameter $\boldsymbol{\theta}$, error variance $\sigma_z^2$ and shape parameter $\nu$. $\nu$ is constrained between $[0.1,3]$.

\item[RLS] AR with Recursive Least Square estimation.
The initial training is the same as AR-$L_\nu$, $P$ is initialised by $\mathbf{X}_B^{}\mathbf{X}_B^\top$, where $\mathbf{X}_B$ is the initial training data.
The adaptive step is implemented following Section \ref{sec:RLS} with $\lambda=0.9999$.
To ensure numerical stability, we only apply the adaptive update if the $L_1$ norm of the update $\lVert \boldsymbol{\theta}^\ast - \boldsymbol{\theta} \rVert_1$ is less than $0.1$.

\item[NR] AR with Newton-Raphson estimation.
Estimates from AR-$L_\nu$ on the training data initialise parameters.
Auxiliary parameters in equation \eqref{eq: NR Update step} are initialised by $\mathbf{h}_0=\mathbf{0}$ and $\nabla^2_{w}l_0(w_0)=\mathbf{0}_{p+2}$.

\item[Bayes] AR with Bayesian estimation and fixed shape parameter $\nu$ as described in Section \ref{sec:Bayes}.
Estimates from AR-$L_\nu$ on the training data initialise parameters.
For the precision of the parameter $\boldsymbol{\theta}$, set $P:= 0.0001 {\cdot} I_{p+1}$.
For the inverse of model error variance $\sigma_z^{-2}$, set $\alpha=101$, $\beta=1$.
The decay factors are constant and set to $\lambda_\theta = \lambda_z = 0.995$.
Then the model is trained on the entire training data $\mathbf{X}$. 
The adaptive update is the same as the initial training by replacing the training data with newly received data, see equations \eqref{bayes: precision update} to \eqref{bayes: beta update}.

\item[Bayes-$\nu$] AR with Bayesian estimation and varying shape parameter.
Same initialisation as Bayes method above. 
During the training step, whether the initial training or the adaptive update, $\nu$ will be updated as described in Section \ref{sec:AdaBayes}.
A matrix singularity check is performed  before each Cholesky decomposition and update of the precision matrix at equation \eqref{eq:Cholesky}.
The learning rate is set to $\gamma=0.05$.
\end{description}

\subsection{Continuous Ranked Probability Score and Skill Score}
\label{sec:CRPS}

As we are in the probabilistic forecasting context, commonly used evaluation metrics for point forecasting, such as Mean Absolute Error (MAE) and Mean Square Error (MSE), are not suitable \citep{gneitingEditorialProbabilisticForecasting2008}. 
To evaluate probabilistic forecasts, the Continuous Ranked Probability Score (CRPS) \citep{Gneiting2007}, a generalisation of MAE and the Brier Score, is used.
Consider a single probabilistic forecast $F$, which is the forecasted CDF, and the observed value $y$.
The CRPS is defined as
\begin{equation}
    \text{CRPS}(F,x) := \int_{-\infty}^{+\infty}\{F(z) - \mathbf{1}_{y \ge z} \}^2 \mathrm{d}z
    ,\end{equation}
where $\mathbf{1}_{y \ge z}$ is the indicator function.
In our case, power generation has support $[\epsilon, 1-\epsilon]$, and the integral can be calculated only on that support.
Notice that a point forecast can be viewed as assigning all probability mass to a single value.
So, it has a step function-shaped CDF, and its CRPS equals its MAE.
For the entire (non-empty) probabilistic forecasts $\{F_t\}_{t=1,2,3,\dots,N}$ with the observation $\{y_t\}_{t=1,2,3,\dots,N}$, CRPS is defined as the average over all single forecast CRPS
\begin{equation}
    \overline{\text{CRPS}} := \frac{1}{N}\sum\nolimits_t \text{CRPS}(F_t,x_t)
    .\end{equation}

$\overline{\text{CRPS}}$ may vary between wind farms due to the unique characteristics of each wind farm.
Thus, the Skill Score is commonly used to evaluate method performance in terms of relative improvement against a benchmark and an ideal result \citep{WHEATCROFT2019573}.
With $\overline{\text{CRPS}}_\text{Persistence}$ as the benchmark result, $\overline{\text{CRPS}}_\text{method}$ as the CRPS of the reference method result, $\overline{\text{CRPS}}_\text{ideal}=0$ as the ideal result, the CRPS Skill Score is defined as
\begin{equation}
    \text{Skill}(\text{method}) := \frac{\overline{\text{CRPS}}_\text{Persistence} - \overline{\text{CRPS}}_\text{method}} {\overline{\text{CRPS}}_\text{Persistence} - \overline{\text{CRPS}}_\text{ideal}}
    .\end{equation}

\begin{figure}
    \centering \includegraphics[width=1\linewidth]{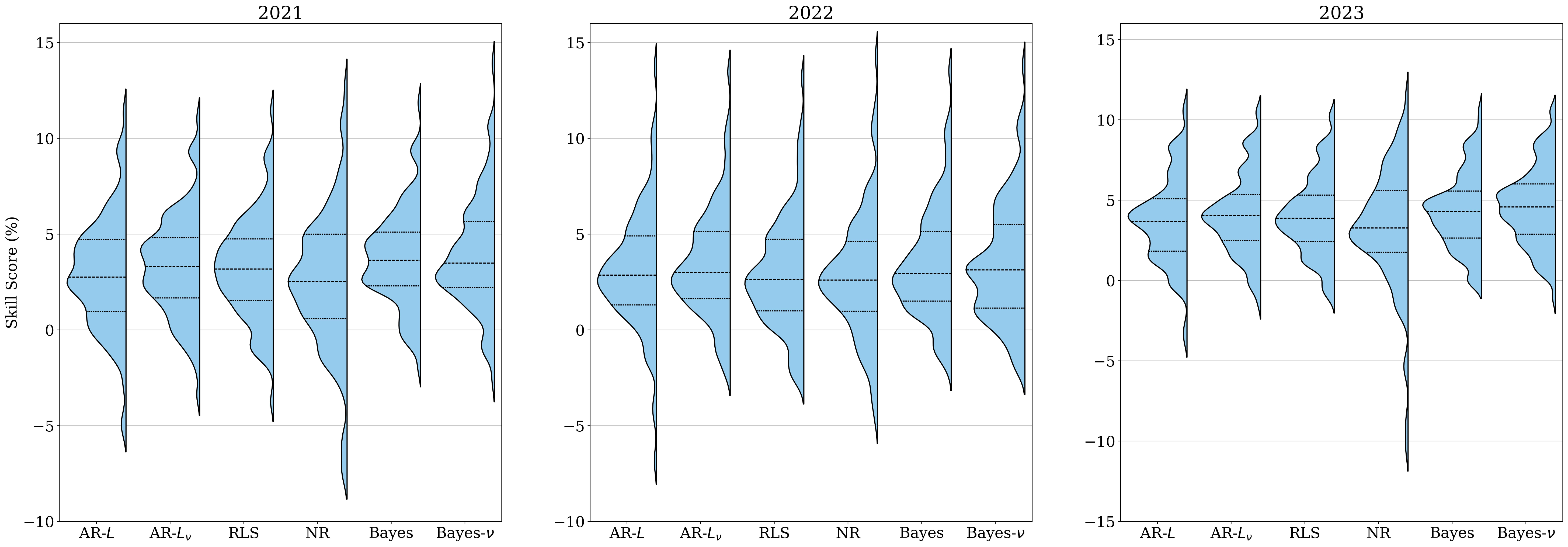}
    \caption{Violin plot of Skill Score relative to Persistence for all three test datasets.}
    \label{fig:Skill}
    \end{figure}

\hfill \break
Table \ref{tb:crps skill} shows the average CRPS and Skill Score over all three test datasets, with the best result for each test dataset highlighted in bold.
For reference, the overall CRPS and Skill Score are summarised in Table \ref{tb:CRPS total}.
To investigate how this improvement is distributed among the 101 wind farms and three test years, Table \ref{tb:rank stats} reports the frequency of each method's rank across all wind farms and test scenarios.
Figure \ref{fig:Skill} provides a more detailed comparison of the Skill Score, by displaying violin plot of the Skill Score of wind farms over three test datasets.

\begin{table}[h!]
    \caption{
    Performance metrics for averaged CRPS(\%) and averaged Skill(\%) over three test datasets.}
    \label{tb:crps skill}
    \centering
    \begin{tabular}{lcccccccc}
    \toprule
    & \multicolumn{3}{c}{\textbf{avg. CRPS (\%)}} &  & \multicolumn{3}{c}{\textbf{avg. Skill (\%)}} \\
    \cline{2-4} \cline{6-8}
    \textbf{Test Dataset} & \textbf{2021} & \textbf{2022} & \textbf{2023} & & \textbf{2021} & \textbf{2022} & \textbf{2023} \\
    \hline
    Persistence           & 3.793 & 4.108 & 3.815 & & 0.000 & 0.000 & 0.000 \\
    AR-$L$                & 3.684 & 3.987 & 3.676 & & 2.975 & 3.106 & 3.751 \\
    AR-$L_\nu$            & 3.671 & 3.976 & 3.664 & & 3.339 & 3.422 & 4.096 \\
    \hline
    RLS                   & 3.676 & 3.996 & 3.667 & & 3.217 & 2.950 & 4.004 \\
    NR                    & 3.701 & 4.001 & 3.694 & & 2.560 & 2.828 & 3.348 \\
    \hline
    Bayes                 & 3.656 & 3.977 & 3.655 & & 3.744 & 3.432 & 4.329 \\
    Bayes-$\nu$           & \textbf{3.651} & \textbf{3.975} & \textbf{3.644} & & \textbf{3.925} & \textbf{3.504} & \textbf{4.604} \\
    \bottomrule
    \end{tabular}
\end{table}

\begin{table}[h!]
    \caption{
    Frequency counts for different methods across ranks on CRPS result over all three test scenarios.
    Rank 1 counts the frequency that the method gets the best CRPS result.
    Rank 7 counts that of the worst CRPS result.
    For Rank 1 to 3, the best result is highlighted in bold.
    }
    \label{tb:rank stats}
    \centering
    \begin{tabular}{lccccccc}
    \toprule
    \textbf{Method} & \textbf{Rank 1} & \textbf{Rank 2} & \textbf{Rank 3} & \textbf{Rank 4} & \textbf{Rank 5} & \textbf{Rank 6} & \textbf{Rank 7} \\
    \hline
    Persistence      & 13 & 7  & 3  & 4  & 6  & 32  & 238 \\
    AR-$L$           & 4  & 41 & 48 & 52 & 95 & 43  & 20 \\
    AR-$L_\nu$       & 31 & 44 & \textbf{78} & 80 & 54 & 16  & 0 \\
    \hline
    RLS              & 1  & 7  & 51 & 73 & 85 & 84  & 2 \\
    NR               & 91 & 37 & 23 & 14 & 12 & 87  & 39 \\
    \hline
    Bayes            & 34 & \textbf{118} & \textbf{74} & 49 & 23 & 5  & 0 \\
    Bayes-$\nu$      & \textbf{129} & 49 & 26 & 31 & 28 & 36 & 4 \\
    \bottomrule
    \end{tabular}
\end{table}

According to Table \ref{tb:crps skill}, Bayes-$\nu$ achieves the lowest average CRPS and the highest Skill Score across all three test datasets.
However, there is variation in skill score between wind farms, illustrated in Table \ref{tb:rank stats} and Figure \ref{fig:Skill}.
Across all test scenarios and wind farms, Bayes-$\nu$ attains Rank 1 a total of 129 times, achieving the lowest CRPS (and consequently the highest Skill Score).
NR attains Rank 1 on 91 occasions but is ranked 6 or 7 a total of 128 times due to high CRPS values.
Bayes demonstrates consistent performance, frequently ranking second or third.
In the 2021 and 2022 test datasets, RLS, NR, Bayes, and Bayes-$\nu$ exhibit a multi-peaked Skill Score density.
Negative Skill Scores are also observed in the violin plots.
The subsequent discussion provides a detailed analysis of these findings.


Firstly, Bayes and Bayes-$\nu$ demonstrate more resilient Skill Score results. 
In the rank statistics (Table \ref{tb:rank stats}), those methods dominate the top 2 rank while maintaining a low record in the bottom two ranks.
In the violin plots (Figure \ref{fig:Skill}), both methods have either a higher lower bound in the Skill Score density or a lower density at lower Skill Scores.
The peaks of the Skill Score density show a slight shift towards higher Skill Scores.

On the other hand, while NR frequently attains Rank 1, it also frequently attains Ranks 6 and 7.
This is a result of instability in the numerical method based on single data updates at some wind farms.
In practical applications, this fragility introduces additional uncertainty, which may hinder decision-making.
The comparison between Bayesian methods and NR suggests that Bayesian methods enhance forecasting performance in a more consistent manner.

Secondly, using generalised logit transformation instead of the standard Logit Transformation can improve forecasting performance, and an adaptive update on shape parameter estimation can further enhance it.
AR-$L_\nu$ and Bayes-$\nu$ achieve a higher average Skill Score across the three test datasets compared to their counterparts with fixed parameters.
This provides evidence of higher accuracy when using the generalised logit transformation.
The differences in average Skill Scores across the three test datasets between Bayes and AR-$L_\nu$
(respectively $0.41\%, 0.01\%, 0.23\%$) are both greater and smaller then that between Bayes-$\nu$ and Bayes (respectively $0.18\%,0.07\%, 0.27\%$).
This verifies the implication stated in Section \ref{sec: DQA}, namely that the improvement in forecasting performance from adaptive estimation of the shape parameter depends on how the shape parameter behaves in reality.
If the shape parameter has slower and more stable variation, the Bayesian component contributes the most to performance enhancement.
If the shape parameter varies more rapidly, adaptive estimation can further enhance performance.

A final remark is that missing data and data on or close to boundaries may contribute to the low Skill Score in some cases.
As shown in the previous analysis, missing data could imply operational discrepancies, regardless of the specific cause.
Such discrepancies may further impact the power generation process and, ultimately, degrade the overall quality of the training and test data.
Adaptive methods track slow changes well but take time to adjust to abrupt changes.
Boundary values may violate the assumption of a Gaussian distribution (even in the transformed domain due to truncation), potentially reducing the effectiveness of adaptive updates near boundary data.
This highlights the critical role of data quality and pre-processing in the practical performance of these methods.
Furthermore, developing more sophisticated methods to handle boundary data and abrupt changes remains a key direction for future research.

\subsection{Reliability Diagram }
\label{sec: reliability}

\begin{figure}
    \centering
    \includegraphics[width=1\linewidth]{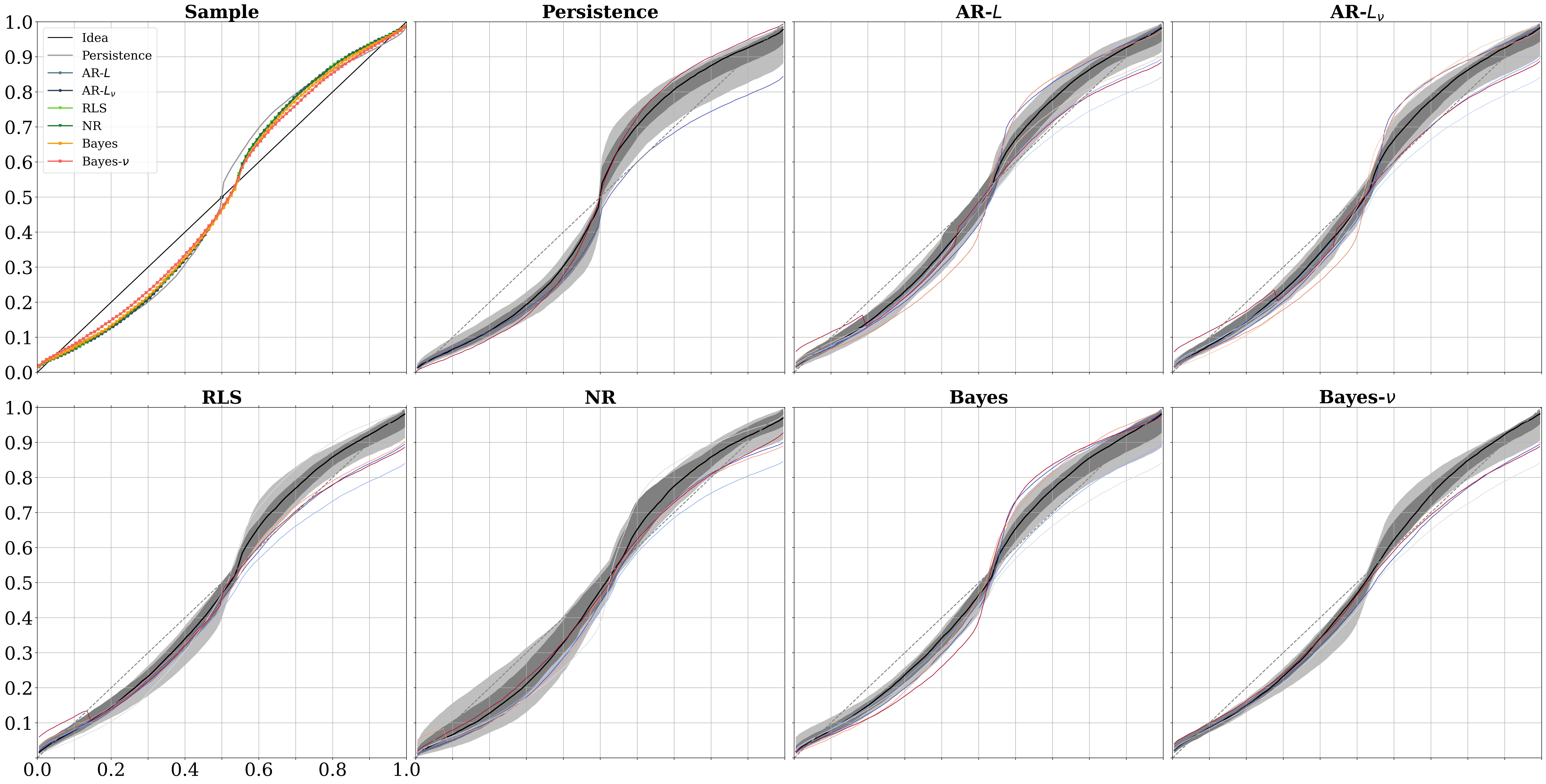}
    \caption{Reliability Diagram and Functional Reliability Diagram \citep{Sun01012011,Hyndman01012010}. 
    The top-left chart provides a comparison of all seven methods for one wind farm, with the x-axis representing the nominal proportion (forecasted CDF) and the y-axis representing the estimated observed proportion (estimated observed CDF).
    The remaining charts are functional reliability diagrams for each method across all wind farms.
    The diagonal line represents the ideal result for reference.
    The dark line represents the median curve across all reliability curves.
    The dark gray envelope indicates the 50\% central region.
    The light gray region illustrates the maximum non-outlying envelope.
    The red to blue single curves are examples of outliers.}
    \label{fig:reliability}
    \end{figure}

Reliability diagrams evaluate the calibration of probabilistic forecasts by illustrating how predicted probabilities correspond to observed outcome frequencies. 
They are presented as P-P plots, offering insights into the alignment between forecasted and actual probabilities \citep{GneitingRegressiondiagnostics2023}.
The top left corner of Figure \ref{fig:reliability} provides an example of the reliability diagram (with cumulative probabilities ranging from $0.5\%$ to $99.5\%$) for all seven methods for one test year for a single wind farm.
The remaining plots in Figure~\ref{fig:reliability} show functional reliability diagrams for each method across all wind farms, constructed using the functional boxplots of \cite{Sun01012011}.
Functional boxplots extend traditional boxplots to functional data, summarising distributions of curves. 
They display the central (median) function, an envelope for the interquartile range, and outliers, aiding in visualising variability and detecting anomalies in functional datasets.
Functional reliability diagrams provide a comprehensive view of forecast performance across multiple wind farms, effectively summarising the calibration of each method.

We can see in Figure~\ref{fig:reliability} that all methods exhibit an ``S''-shaped curve, indicating that they tend to produce over-dispersed predictive densities---meaning they estimate greater uncertainty than the ground truth. 
However, the two Bayesian methods demonstrate slightly better performance compared to the others.
Notably, the width of $50\%$ region for RLS and NR methods is larger than that for Bayes and Bayes-$\nu$ indicating a greater variation in bias, and calibration more general, between wind farms.
This is another reflection of higher CRPS of RLS ans NR methods compared to the Bayes methods.

\subsection{Single Wind Farm Example} \label{sec:swfe}

\begin{figure}
    \centering
    \includegraphics[height=0.9\textheight]{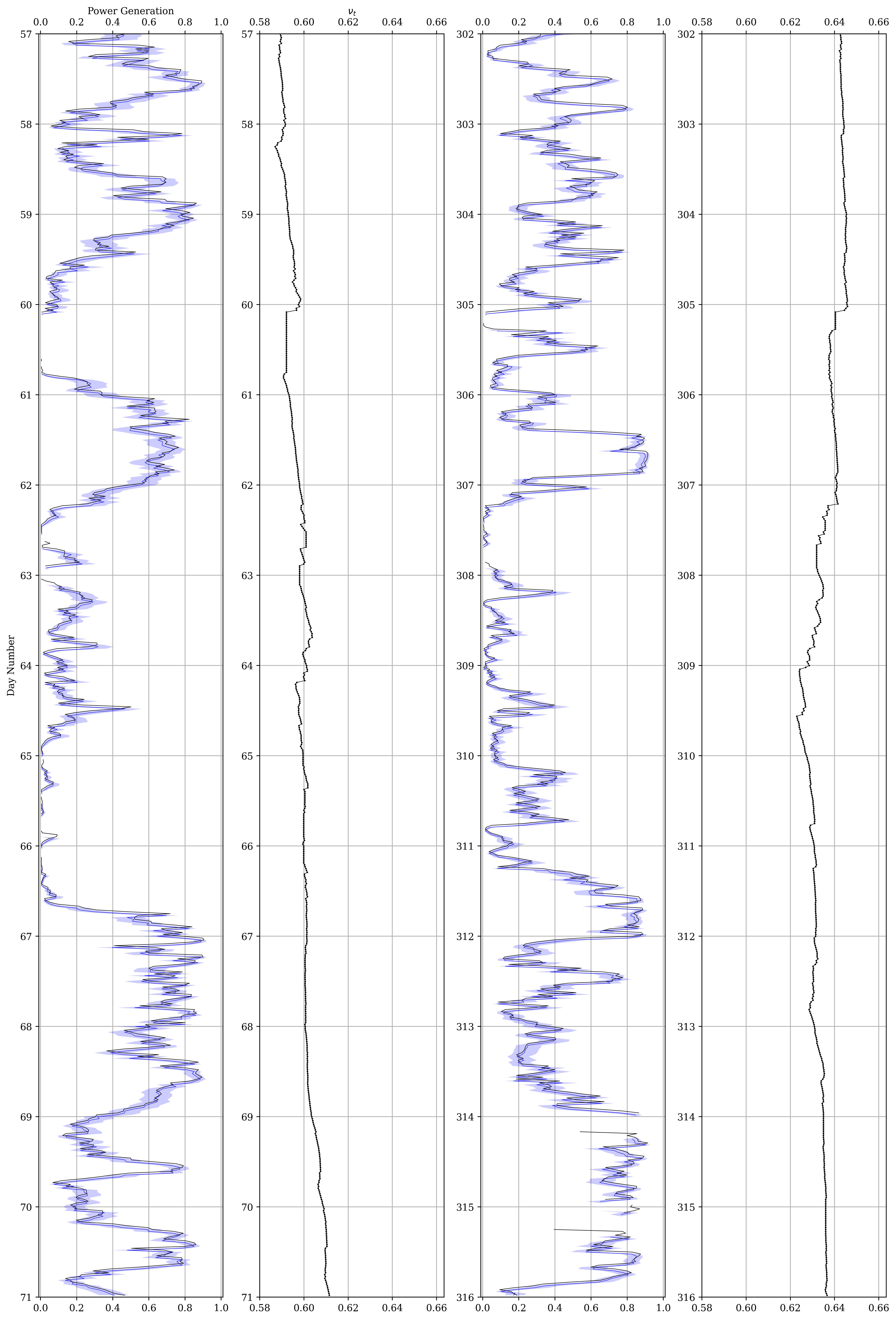}
    \caption{
    A case study of probabilistic forecasts is presented for a single wind farm using the Bayes-$\nu$ method.
    Forecasts spanning two fortnights, indexed by day number, are illustrated.
    The observed data is represented by a dark line, the $50\%$ quantile forecasts by a blue line, and the forecast interval, covering the 25\% to 75\% quantiles, is depicted as a blue patch.
    The variation of $\nu_t$ is shown alongside the forecast plot. }
    \label{fig:case study}
    \end{figure}

To offer direct visual insight, Figure \ref{fig:case study} presents a case study of a single wind farm.
Given the large volume of data, we selected one month's worth for readability. 
The chosen period minimises missing values while capturing a broad range of observed power generation. 
One dataset begins in mid-February, and the other starts in October.

We can see from figure \ref{fig:case study} that the forecast uncertainty (represented by the blue shading) becomes narrower when the power generation is near the bounds (both lower and upper).
For example, compare the probabilistic forecasts where the power generation is within the range $(0,0.2) \cup (0.8, 1)$ with those where the power generation is within the range $(0.2, 0.8)$. 
The width between the $75\%$ and $25\%$ quantiles, i.e., the width of the blue shading, for the former is typically less than $0.1$.
In contrast, the width of the latter can reach $0.2$, which is twice that of the former.
This behaviour of the forecasted uncertainty confirms the finding regarding subsequent power generation based on previous observations, where the variance near the bounds is smaller \citep{pinsonAdaptiveModellingforecasting2012}.

With regard to the shape parameter, it shows a mild decrease, increase, and levelling off during the presented periods.
We also observed surges and declines in the shape parameter in other cases.
Regardless of how the shape parameter varies, there is no obvious correlation with power generation.
Therefore, there is potential to use the time series of the shape parameter to perform real-time anomaly analysis when needed.

\subsection{Sensitivity Analysis} \label{sec:Sensitivity Analysis}

\begin{figure}
    \centering\includegraphics[width=1\linewidth]{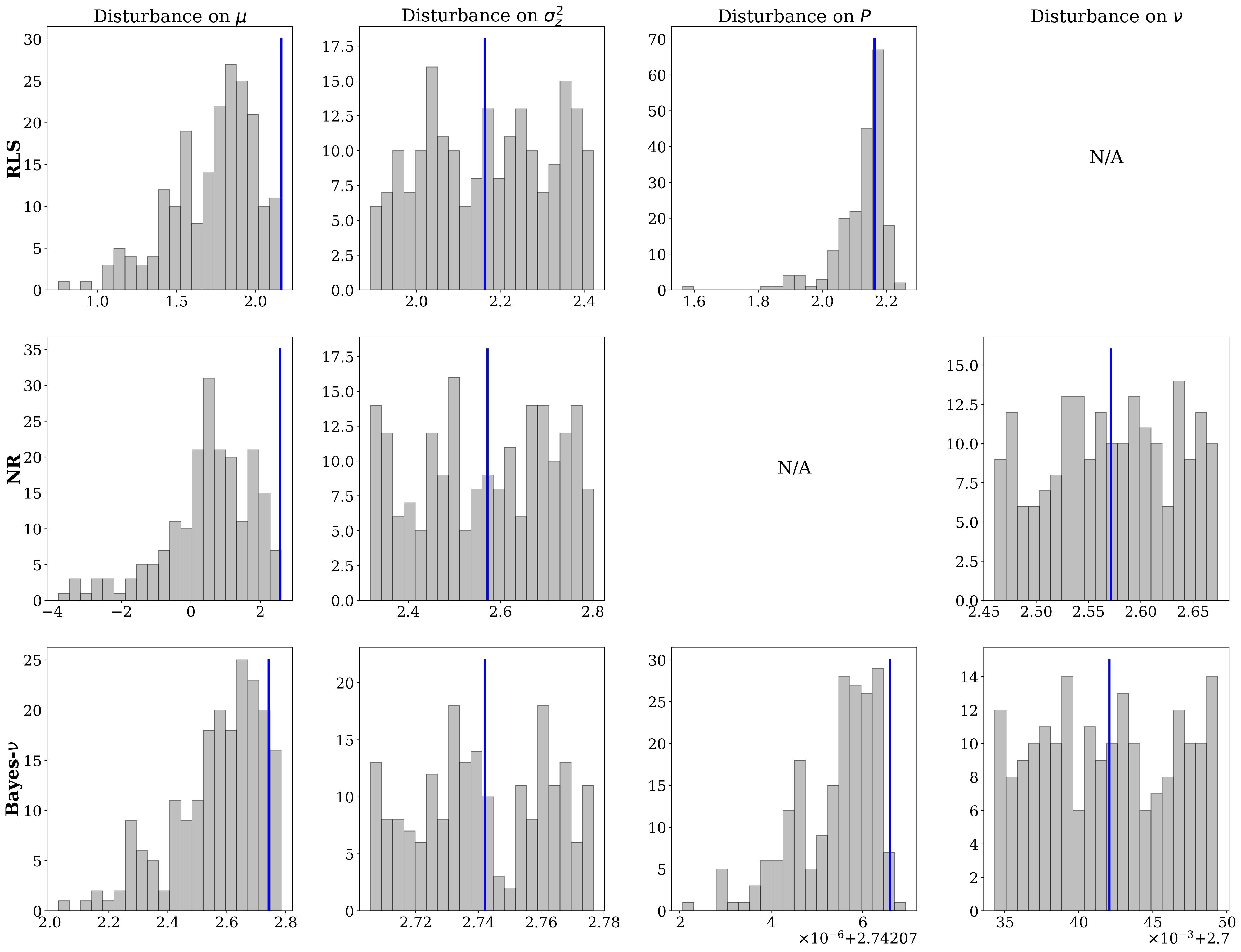}
    \caption{Sensitivity Analysis across the RLS, NR and Bayes-$\nu$ methods.
    Histogram of Skill Score are used to evaluate the robustness of the methods.
    The x-axis represents the range of Skill Scores in percentage ($\%$).
    It is important to note that the scale and range of the x-axis vary depending on the method and the type of disturbance applied.
    The y-axis is the frequency of tested result corresponding to each bin.
    The blue stick represents the original Skill Score without any disturbance.
    }
    \label{fig:sensitivity}
    \end{figure}

In adaptive forecasting, robustness is an essential requirement \citep{pmlr-v151-yoon22a}. 
Therefore, we conducted a sensitivity analysis on the RLS, NR, and Bayes-$\nu$ methods to evaluate their robustness against disturbances.
We selected a wind farm where the three methods demonstrated comparable Skill Scores.
At the end of the training, we introduced disturbances to the model parameters $\mu$ and the model variance $\sigma_z^2$ for all three models.
 We also introduced disturbances to the intermediate matrix $P$ in the update equations \eqref{eq: P smoothing} and \eqref{bayes: precision update} for the RLS and Bayes-$\nu$ methods, as well as to the shape parameter $\mu$ for the NR and Bayes-$\nu$ methods.
Figure \ref{fig:sensitivity} presents the Skill Score density from 200 random simulation results.

For all methods, disturbances to the model parameters $\mu$ and the matrix $P$ (if applicable) result in decreased performance.
This indicates that the method indeed finds a locally optimal estimation.
 When the model variance $\sigma_z^2$ and the trained shape parameter $\nu$ are disturbed, all methods exhibit an evenly distributed performance discrepancy centred around the original Skill Score.
This suggests that the trained model variance and shape parameter are not locally optimal estimations for the test data, and that adjusting these parameters could enhance model performance.
Thus, this case confirms the practical reality that the shape parameter varies in real-world wind power forecasting applications and that adaptive methods should be employed by forecasters.

The Bayes-$\nu$ method demonstrates highest robustness compared to the other two methods.
Firstly, under disturbances to the model parameters $\mu$, the RLS and NR methods experience a severe decline.
The Skill Score of RLS drops from $2.2\%$ to $0.8\%$, while the Skill Score of NR even falls below $0\%$.
This indicates that when an incident distorts the model, its performance degrades to a level similar to the naive Persistence method.
Similar results are observed for disturbances to the model variance, $P$ matrix, and shape parameter.
This highlights a potential drawback of the RLS and NR methods, which update directly based on forecast discrepancies or first-order derivatives and may face numerical stability issues.
For the Bayes-$\nu$ method, the impact of disturbances on the $P$ matrix and the shape parameter $\nu$ is extremely small, at levels of $10^{-6}\%$ and $10^{-3}\%$, respectively, while the impact on model variance is limited to approximately $0.1\%$.
This result demonstrates a strong self-adaptation capability of the Bayes-$\nu$ method.
The observed robustness can be attributed to its step-wise and interactive update mechanism.
For example, the $P$ matrix is not only utilised for model parameter updates but also reconstructs representative data to adaptively update the shape parameter.
Consequently, disturbances to $\nu$ can be quickly corrected by the term ${L}^\ast(\nu)$ in Equation \eqref{eq:bayes nu}.

\section{Conclusion}\label{sec:cls}

This paper introduces an adaptive probabilistic forecasting method that integrates the generalised logit transformation with online (adaptive) Bayesian estimation, addressing non-stationarity in very short-term forecasting of wind power production.
The generalised logit transformation maps data from a double-bounded support to an unbounded domain, enabling the application of a Gaussian assumption in the transformed domain.
Bayesian estimation offers improved forecast performance and robustness in the estimation of autoregressive-type models, and adaptive updates in this setting are shown to be more robust than other numerical methods.
In particular, we have proposed a novel adaptive update mechanism for the shape parameter, leveraging Bayesian updates to construct representative data and optimise the shape parameter via a dedicated objective function.

Seven methods were evaluated using wind farm data collected from over 100 wind farms over four years.
The model employing the Bayesian method with an adaptive shape parameter demonstrated consistent improvements in both CRPS and reliability.
Further analysis also confirmed the robustness that the Bayesian method can provide.
We also observed that, in probabilistic forecasting, numerical stability and uncertainty estimation are crucial factors affecting performance.
Moreover, an appropriately estimated uncertainty is necessary for stable performance.

Wind power forecasting is a prime example of a complex scenario where a probabilistic approach has practical value for end-users.
An adaptive and probabilistic forecasting approach based on the Bayesian method can react to streaming data in real time and ensure consistent processing.
Future research should explore the integration of this flexible updating capability into more complex, physics-based models, as well as the handling of changing boundaries to improve the data processing step.

\section*{Acknowledgements}\label{sec:ackn}

The authors appreciate the financial support provided by the China Scholarship Council.
The authors appreciate the access to wind farm data provided by Elexon, this analysis contains BSC information licensed under the BSC Open Data Licence (https://www.elexon.co.uk/bsc/data/open-data-requests/bsc-open-data-licence/).


\appendix
\renewcommand{\thefigure}{\Alph{figure}}  
\renewcommand{\thetable}{\Alph{table}}    
\setcounter{figure}{0}  
\setcounter{table}{0}   

\section{CRPS and Skill Score Result Over All Wind Farms}
\label{apd:CRPS}

Table \ref{apd:CRPS} shows the CRPS and Skill Score result for all wind farms including outlier wind farms, those with significant changes in available capacity.
The comparison between this table and Table \ref{tb:crps skill} indicates the same ranking of methods is observed, and the adaptive methods, especially the proposed methods, have even greater skill relative to static methods.
However, as discussed in Section \ref{sec: DQA} and \ref{sec:CRPS}, those outlier wind farms may need more further specific pre-processing in order to get meaningful forecasting results.

\begin{table}[h!]
    \caption{Performance metrics for averaged CRPS(\%) and averaged Skill(\%) over three test datasets, for all wind farms including outlier wind farms.}
    \label{tb:CRPS total}
    \centering
    \begin{tabular}{lcccccccc}
    \toprule
    & \multicolumn{3}{c}{\textbf{avg. CRPS (\%)}} &  & \multicolumn{3}{c}{\textbf{avg. Skill (\%)}} \\
    \cline{2-4} \cline{6-8}
    \textbf{Test Dataset} & \textbf{2021} & \textbf{2022} & \textbf{2023} & & \textbf{2021} & \textbf{2022} & \textbf{2023} \\
    \hline
    Persistence           & 3.676 & 3.971 & 3.697 & & 0.000 & 0.000 & 0.000 \\
    AR-$L$                & 3.567 & 3.845 & 3.546 & & 3.008 & 3.362 & 4.460 \\
    AR-$L_\nu$            & 3.554 & 3.832 & 3.532 & & 3.375 & 3.714 & 4.834 \\
    \hline
    RLS                   & 3.559 & 3.847 & 3.536 & & 3.264 & 3.595 & 4.729 \\
    NR                    & 3.579 & 3.857 & 3.564 & & 2.747 & 3.085 & 4.008 \\
    \hline
    Bayes                 & 3.538 & 3.824 & 3.524 & & 3.842 & 4.352 & 5.041 \\
    Bayes-$\nu$           & \textbf{3.529} & \textbf{3.822} & \textbf{3.518} & & \textbf{4.172} & \textbf{4.450} & \textbf{5.206} \\
    \bottomrule
    \end{tabular}
\end{table}

\hfill \break
\hfill \break
\hfill \break
\hfill \break
\hfill \break

\baselineskip 16pt
\bibliographystyle{CUP}
\bibliography{reference.bib}

\end{document}